\documentstyle[aps,prb]{revtex}
\begin{document}
\draft
\twocolumn[\hsize\textwidth\columnwidth\hsize\csname @twocolumnfalse\endcsname

\title{On induced $CPT$-odd Chern-Simons terms in 3+1 effective
action. }
\author{G.E. Volovik\\
Helsinki University of Technology,
Low Temperature Laboratory,\\
P.O.Box 2200,
FIN-02015 HUT,  Finland, \\
 Landau Institute for Theoretical Physics,
117334 Moscow,
Russia }

\date{\today}

\maketitle

\begin{abstract}
This paper was originally designated as Comment to the paper by R.
Jackiw and
V.  Alan  Kosteleck$\rm{\acute y}$ \cite{JackiwKostelecky}.  We
provide an
example of the fermionic system, the superfluid $^3$He-A, in which the
$CPT$-odd Chern-Simons terms in the
effective action are unambiguously induced by chiral fermions.
  In this system the Lorentz and gauge
invariances both are violated at high energy, but the behavior of the
system
beyond the cut-off is known.  This allows us to construct the $CPT$-
odd
action, which combines the conventional 3+1
Chern-Simons term and the mixed axial-gravitational Chern-Simons term
discussed in Ref.\cite{VolVil}.  The influence of Chern-Simons term
on the
dynamics of the effective gauge field has been experimentally
observed in
rotating $^3$He-A.
\end{abstract}
\


\

]
\narrowtext

Recently the problem of the radiatively induced $CPT$-odd Chern-
Simons term
in 3+1 quantum field theory has been addressed in a number of papers
\cite{ColladayKostelecky,JackiwKostelecky,Fosco,Andrianov,Chung,Perez}.
The  Chern-Simons (CS) term
$ L_{\rm CS} = {1\over 2}k_\mu e^{\mu\alpha\beta\gamma}F_{\alpha\beta
}A_\gamma$
 in the 3+1 electromagnetic action,
where $k^\mu$ is a constant 4-vector, is induced by the $CPT$- and
Lorentz-
violating axial-vector term $b^\mu\gamma_\mu \gamma_5$ in the Dirac
Lagrangian for massive fermions.  In the limit of small and large $b$
compared with mass $m$ of Dirac fermions, it was found that
\begin{eqnarray}
k^\mu= {3\over 16\pi^2} b^\mu ~~,~~b\ll m~, \label{1}\\
k^\mu=- {1\over 16\pi^2} b^\mu ~~,~~b\gg m~. \label{2}
 \end{eqnarray}
However it has been concluded that the existence of CS term  depends
on the
choice of regularization procedure -- "a renormalization
ambiguity".  This means that the result for $k_\mu$ depends on
physics beyond
the cut-off.

The above $CPT$-odd term can result not only from the violation of
the $CPT$
symmetry in the vacuum. The nonzero density of the chiral fermions
violates
the $CPT$ invariance and thus can also lead to the Chern-Simons term,
with
$b^0$ determined by the chemical potential $\mu$ and temperature $T$
of the fermionic system \cite{Vilenkin,Redlich,JoyceShaposhnikov}.

Here we provide an example of the fermionic system, in which such
Chern-Simons term is unambiguously induced by fermions.
In this system the Lorentz and
gauge invariances both are violated at high energy, but the behavior
of the
system beyond the cut-off is known. This allows the calculation of CS
term in
different physical situations. The influence of this CS term on the
dynamics
of the effective gauge field has been experimentally observed.

This is the superfluid
$^3$He-A, where in the low-energy corner
there are two species of fermionic quasiparticles:
left-handed and right-handed Weyl fermions \cite{LammiTalk}.
Quasiparticles
interact with the order parameter, the unit $\hat{\bf l}$-vector of
the
orbital momentum of Cooper pairs, in the same manner as chiral
relativistic
fermions interact with the vector
potential of the $U(1)$ gauge field: ${\bf A}\equiv p_F\hat{\bf l}$,
where
$p_F$ is the Fermi momentum.  The "electric charges" -- the charges
of the
left and right quasiparticles with respect to this effective gauge
field --
are $e_R=-e_L=-1$.  The normal component
of superfluid $^3$He-A consists of the thermal fermions, whose density
is determined by $T$ and by the velocity ${\bf v}_n-{\bf v}_s$
of the flow of the
normal component with respect to the superfluid
vacuum. The velocity of the counterflow in the direction of
$\hat{\bf l}$ is equivalent to the chemical potentials
for left and right fermions in relativistic system:
\begin{equation}
\mu_R=-\mu_L= p_F\hat{\bf l}\cdot({\bf v}_n-{\bf v}_s) ~.
\label{ChemicalPot}
\end{equation}

As in the relativistic theories, the state of the system of
chiral quasiparticles with nonzero counterflow velocity (an analogue
of
chemical potential)
violates Lorentz invariance and $CPT$ symmetry and induces the $CPT$-
odd
Chern-Simons term.  This term can be written in general form, which
is valid
both for the relativistic systems including that found in
Ref.\cite{Vilenkin,Redlich} and for $^3$He-A \cite{LammiTalk}:
 \begin{equation}
{1\over
4\pi^2}\left(\sum_L \mu_L e^2_L-\sum_R \mu_R e^2_R\right) {\bf
A}\cdot(\vec\nabla\times {\bf A})~. \label{ChernSimonsGauge}
 \end{equation}
Here sums over $L$ and $R$ mean summation over all the left-handed and
right-handed fermionic species respectively; $e_L$ and  $e_R$ are
charges of
left and right fermions with respect to $U(1)$ field (say,
hypercharge field
in the Standard model).

Translation of Eq.(\ref{ChernSimonsGauge}) to the $^3$He-A language
gives
 \begin{equation}
 {p_F^3\over 2\pi^2}\left(\hat{\bf l}_0\cdot({\bf v}_s-{\bf v}_n)
\right)
 \left(\delta\hat{\bf l}\cdot(\vec\nabla\times \delta\hat{\bf
l})\right)~.
 \label{ChernSimonsHe}
 \end{equation}
 Here $\hat{\bf l}_0$ is the direction of the order parameter
 $\hat{\bf l}$ in the
 homogeneous ground state; $ {\bf v}_n-{\bf v}_s$ is the uniform
 counterflow of the fermionic quasiparticles with respect to
 the superfluid vacuum; and $\delta \hat{\bf l}=\hat{\bf l} -
\hat{\bf l}_0$
 is the deviation of the order parameter from its ground state
direction.

 Since for chiral fermions the chemical potential plays the part
 of the parameter $b^0$ in the fermionic Lagrangian,
 the connection between $k^0$ and $b^0$ is
$k^0= b^0/ 2\pi^2$ in $^3$He-A. Though it agrees with
the result obtained in relativistic system with nonzero chemical
potential
for chiral fermions \cite{Vilenkin}, it does not coincide
with Eq.(\ref{2}) obtained in the massless limit $ m/b^0\rightarrow
0$.

The instability of the electromagnetic vacuum due to the 3+1 Chern-
Simons
term has been discussed by Caroll, Field
and Jackiw \cite{CarollFieldJackiw}, Andrianov and Soldati
\cite{AndrianovSoldati}, and   Joyce and Shaposhnikov
\cite{JoyceShaposhnikov}. In the case of the nonzero density of
 right electrons ($\mu_R\neq 0$) this instability leads to the
conversion
 of the density of the right electrons to
 the hypermagnetic field. This effect was used in the
 scenario for  nucleation of the
primordial magnetic field \cite{JoyceShaposhnikov}.
In  $^3$He-A this phenomenon is represented by the well known
helical instability of the counterflow, which is triggered by the term
in Eq.(\ref{ChernSimonsHe}) \cite{VollhardtWolfle}. The
conversion of the counterflow of the normal component
(an analogue of  $\mu_R$ in the Joyce-Shaposhnikov
scenario) to the inhomogeneous ${\hat{\bf l}}$-field with
$\nabla\times
\hat{\bf l}\neq 0$  (an analogue of
hypermagnetic field)
due to this instability has been observed in rotating
$^3$He-A \cite{Experiment,LammiTalk}.

Recently another type of the Chern-Simons term has been found
for both systems, $^3$He-A and chiral relativistic fermions with
nonzero
$\mu$ or/and $T$. This is the mixed axial-gravitational CS term,
which contains
both the gauge field and the gravimagnetic field \cite{VolVil}:
 \begin{eqnarray}
{1\over 8\pi^2}\left(\sum_L \mu_L^2 e_L-\sum_R \mu_R^2 e_R\right)
{\bf A}\cdot {\bf B}_{\bf g} ~,
\label{ChernSimonsMixed}\\
\nonumber
{\bf B}_{\bf g}=\vec\nabla\times {\bf
g}~,~
{\bf g} \equiv g_{0i} ~.
\end{eqnarray}
Here $g_{0i}$ is the element of the metric in the reference
frame of the heat bath
(in superfluids it is the element of the effective metric in the
frame, in
which the normal component is at rest). If the heat bath
of chiral fermions is rotating in  Minkowski space, the
"gravimagnetic field" is expressed in terms of rotation velocity
${\bf \Omega}$:
\begin{equation}
{\bf B}_{\bf g}=\nabla\times {\bf g} =2{{\bf \Omega}\over c^2} .
\label{GravMagField}
\end{equation}
Here $c$ is the material parameter, which is the speed of
light in relativistic system, and the initial slope
in the energy spectrum of fermionic quasiparticles
propagating in the plane transverse to the $\hat {\bf
l}$-vector in $^3$He-A \cite{LammiTalk}.  The  material parameters
do not enter Eq.(\ref{ChernSimonsMixed}) explicitly: they enter
only through the metric. That is
why the same equation Eq.(\ref{ChernSimonsMixed})
can be applied to different fermionic systems,
including those with varying speed of light.
In relativistic system this equation describes the macroscopic
parity violating effect: rotation of the heat bath (or of the
black hole) produces the flux of the
chiral fermions along the rotation axis \cite{Vilenkin2}.

Comparison of the Eqs.(\ref{ChernSimonsMixed}) and
(\ref{ChernSimonsGauge})
suggests that the two $CPT$-odd terms can be united if one uses the
Larmor theorem and introduces the combined fields:
\begin{equation}
{\bf A}_{L(R)}=e_{L(R)} {\bf A}+{1\over 2} \mu_{L(R)} {\bf g}~,~{\bf
B}_{L(R)}
=\nabla\times {\bf A}_{L(R)} ~.
\label{MagneticGravimagnetic}
\end{equation}
Then the general form of the Chern-Simons $CPT$-odd term is
\begin{equation}
{1\over 4\pi^2}\left(\sum_L \mu_L {\bf A}_L\cdot {\bf B}_L -\sum_R
 \mu_R {\bf A}_R\cdot {\bf B}_R\right)  ~.
\label{ChernSimonsGeneral}
\end{equation}

Note that in the Standard Model
the nullification of the $CPT$-odd term in
Eq.(\ref{ChernSimonsGeneral})
occurs if the "gyromagnetic" ratio $e/\mu$ is the same
for all fermions.
This happens because of the anomaly cancellation. For the
$CPT$-odd term induced by the vacuum fermions, the anomaly
cancellation
was discussed in Refs.\cite{Perez,ColladayKostelecky}. In $^3$He-A
the "gyromagnetic ratio" is the same for two fermionic species,
$e_L/\mu_L=e_R/\mu_R$, but the CS terms survive, since
there is no anomaly cancellation in this system.

In $^3$He-A there are also subtle points related to gauge
 invariance of the CS term, as discussed by Coleman and Glashow
  \cite{ColemanGlashow}, and to the reference frame. They
are  determined by physical situations.

 (i) The reference
frame for superfluid velocity ${\bf v}_s$ is  the heat bath
frame -- the frame of the normal component
moving with velocity ${\bf v}_n$. At $T= 0$
this frame disappears: thermal fermions are frozen out.
To avoid  uncertainty in determination of the counterflow
velocity ${\bf v}_s - {\bf
v}_n$, and thus of the chemical potential of the chiral fermions, the
limit
$T \rightarrow 0$ must be taken after all other limits.

(ii) The leading terms in the low-energy
effective action for the "electrodynamics" of $^3$He-A are gauge
invariant, because the main contributions
to the effective action are induced
by the low-energy fermions, which are "relativistic"
and obey the gauge invariant Lagrangian. The
Eq.(\ref{ChernSimonsGeneral}) is an example of such
gauge invariant term in the low-energy  action.  It is gauge
invariant if the
$b^0$ parameter (or $\mu_R$) is
constant, i.e. if the background counterflow and ${\hat{\bf
l}}_0$ field are homogeneous. The inhomogeneous corrections, which
correspond to the inhomogeneous $b^0$, violate the gauge invariance.
This is natural, since these corrections are determined by the
higher energy fermions, which do not obey the gauge invariance
from the very beginning.
This is in agreement with the
conclusion made in Ref.\cite{JackiwKostelecky}, that for existence
of the CS term the "weak condition" -- the gauge
invariance at zero 4-momentum -- is required.

I thank Alex Vilenkin for discussions. This work was supported in
part by the
Russian Foundations for Fundamental Research grant No. 96-02-16072
and by
European Science Foundation.


\begin{references}

\bibitem{JackiwKostelecky}
R. Jackiw and V. Alan Kosteleck$\rm{\acute y}$, Phys. Rev. Lett.
{\bf 82}, 3572 (1999).

\bibitem{VolVil} G.E. Volovik and A. Vilenkin, hep-ph/9905460.

\bibitem{ColladayKostelecky}
D. Colladay and V. Alan Kosteleck$\rm{\acute y}$, Phys. Rev.
{\bf D~58}, 116002 (1998).

\bibitem{Fosco}  C.D. Fosco and J.C. Le Guillou, hep-th/9904138.

\bibitem{Andrianov} A.A. Andrianov, R. Soldati and L. Sorbo, Phys.
Rev.
{\bf D~59}, 025002/1-13 (1998).

\bibitem{Chung}  J.-M. Chung. hep-th/9905095.

\bibitem{Perez}  M. Perez-Victoria, hep-th/9905061.

\bibitem{Vilenkin} A. Vilenkin,
Phys. Rev.  {\bf D~ 22}, 3080  (1980).

\bibitem{Redlich} A.N. Redlich and L.C.R.
Wijewardhana,  Phys. Rev. Lett. {\bf 54}, 970  (1985).

\bibitem{JoyceShaposhnikov}  M. Joyce, M. Shaposhnikov,  Phys.
Rev. Lett. {\bf 79}, 1193  (1997).

\bibitem{LammiTalk} G.E. Volovik,  Physica {\bf B~255}, 86 (1998).

\bibitem{CarollFieldJackiw} S.M. Caroll, G.B. Field
and R. Jackiw, Phys. Rev.
{\bf D~41}, 1231 (1990).

\bibitem{AndrianovSoldati} A.A. Andrianov and R. Soldati, Phys. Lett.
{\bf B~435}, 449 (1998).

\bibitem{VollhardtWolfle} D. Vollhardt, P. and P. W\"olfle,  The
superfluid phases of helium 3,  Taylor and Francis, London - New York
-
Philadelphia, 1990.

\bibitem{Experiment} V.M.H. Ruutu, J. Kopu, M. Krusius, U. Parts, B.
Pla\c{c}ais, E.V. Thuneberg, and W. Xu,
Phys. Rev. Lett. {\bf 79}, 5058 (1997).

\bibitem{Vilenkin2} A. Vilenkin,
Phys. Rev.  {\bf D~ 20}, 1807  (1979);
Phys. Rev.  {\bf D~ 21}, 2260  (1980).


\bibitem{ColemanGlashow} S. Coleman and S. Glashow, hep-ph/9812418.

\end{references}
\end{document}